\documentclass[twocolumn,showpacs,prb]{revtex4}
\usepackage{amsfonts}
\usepackage{amsmath}
\usepackage{amssymb}
\usepackage{graphicx}
\usepackage{natbib}
\usepackage{color}
\usepackage[colorlinks=true, letterpaper=true, pdfstartview=FitV, linkcolor=blue, citecolor=blue, urlcolor=blue]{hyperref}

\begin{document}

\title{Tunable anisotropic behaviors in phosphorene under periodic potentials in arbitrary directions}
\author{Xiaojing Li,$^1\dag$ Wen Yang,$^2$ Kun Luo,$^3$ and Zhenhua Wu$^3\ddag$ \footnote{$^{\dag}$xjli@fjnu.edu.cn\\$^{\ddag}$wuzhenhua@ime.ac.cn.}}
\affiliation{$^{1}$College of Physics and Energy, Fujian Normal University, Fuzhou
350117, China}
\affiliation{$^{2}$Beijing Computational Science Research Center, Beijing 100094, People.s Republic of China}
\affiliation{$^{3}$Key Laboratory of Microelectronic Devices and Integrated Technology,
Institute of Microelectronics, Chinese Academy of Sciences, Beijing 100029,
P. R. China}

\pacs{68.65.Hb, 71.35.Ji, 78.20.Ls}

\begin{abstract}
We investigate theoretically the anisotropic electronic and optical behaviors of a monolayer black phosphorus
(phosphorene) modulated by periodic potential superlattices in arbitrary directions.
We demonstrate that different strength and orientation of the phosphorene potential superlattice can give rise to distinct energy spectra,
i.e., tuning the intrinsic electronic anisotropy. Accordingly, the anisotropic effective mass, and optical absorption modulated by superlattice strength and orientation are addressed systematically. This feature enables tuning capability more than one order of magnitude in the optical absorption spectrum.
Our findings should be useful in building phosphorene optical and (opto)electronic devices by applying external potential superlattice.
\end{abstract}

\maketitle

\section{Introduction}

The enormous efforts devoted to the study of two dimensional (2D) semiconductor materials in the past few years have uncovered numerous extraordinary properties, including tunable bandgap, high carrier mobility, excellent gate electrostatics, offering exciting opportunities for development of high performance electronic and optical devices.
Among them, phosphorene with anisotropic rectangular lattice exhibits the $C^{1}_{2h}$ subgroup, that develops unique highly anisotropic energy dispersions unlike other 2D materials~\cite{Dresselhaus,szhang,shiying,Andrey,Andrey2,Bo}. Recent experiments have also confirmed the theory with observing a strongly anisotropic conducting behavior \cite{QIAO}, anisotropic exciton~\cite{,VyTran,XWang} or optical response~\cite{Xia,YXie,xiao2}, anisotropic Landau levels~\cite{XZhou4} and anisotropic Rashba spin-orbit coupling~\cite{Popovic}.

Furthermore, modulation of the anisotropy by external fields has been successfully proposed and attracted increasing research interests.
People have demonstrated effective band engineering and optical modulation by the real superlattice structure as well as artificial superlattice, e.g., periodic external fields. The Superlattices lead to interesting phenomena such as negative differential conductivity and gap openings at the mini-Brillouin-zone boundary and so on.
In these previous studies, spatially periodic electric and/or magnetic fields are applied to graphene.~\cite{C,Park,Brey,Barbier,ZWu1,Maksimova,ZWu,Uddin,zhengli,ZWu3,Anna0,Le,Anna,Uddin2,XLi}
This basic modulation mechanism by superlattice is also applicable for phosphorene. The effects of magnetic fields on phosphorene's electronic and optical properties have been studied.~\cite{Ono,XZhou2,XZhou3,LLi,RZhang} The magnetic fields break the time-reversal symmetry and develop an in-line anisotropy along the zigzag or armchair direction. However, compared with the magnetic control device, an electronic control device is easier to realize, since one can hardly generate periodic magnetic field in nanoscale integrated devices. More importantly, the physics of the coupling between anisotropic external field and the intrinsic anisotropy in pristine phosphorene has not been studied thoroughly. The investigation of a twisted 2D structure~\cite{PKang,YCao} of phosphorene and periodic external field superlattice in arbitrary direction rather than that only along the zigzag/armchair direction is also highly desired to illustrate the corresponding orientation dependant impacts on the electronic and optical properties. It may enable potential phosphorene optical and (opto)electronic devices design or development of novel applications.

\begin{figure}[tbp]
\includegraphics[width=\columnwidth]{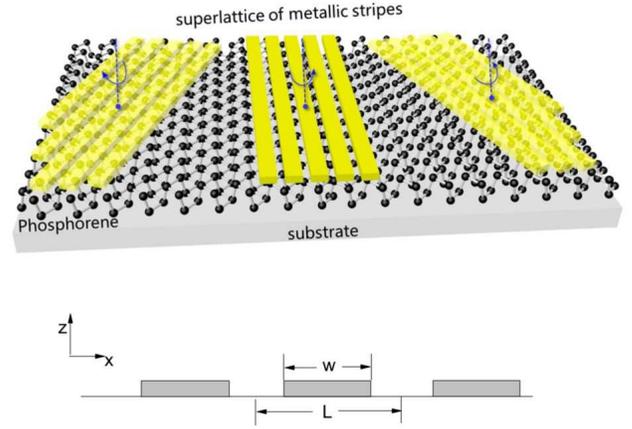}
\caption{Schematic diagram of a phosphorene with periodic metallic stripes of arbitrary twisted angle (upper), $L$ is the superlattice cell length, $W$ is the potential stripe width along the periodic direction (lower).}
\label{fig1}
\end{figure}

In this work, we propose a twisted 2D artificial superlattice composed by phosphorene and
periodic rectangular metal stripes covered above to provide periodic potential, which is an ideal system to archive anisotropic two-dimensional electron system with high flexibility and controllability. The materials of the substrate can be $SiO_{2}$, $Al_{2}O_{3}$, $HfO_{2}$ in real system. We show that the
anisotropic energy dispersions of the phosphorene can be effectively tuned
by different the strength and orientations of the potential fields. Accordingly we illustrate the
impacts of such a periodic field on the anisotropic effective mass and orientation-dependent optical absorption.

\section{Theoretical model}

We consider phosphorene coated by periodic potential as shown in Fig.~\ref{fig1}. The low-energy dispersion of phosphorene can be well described
by a two band $\vec{k}\cdot\vec{p}$ effective mass Hamiltonian due to the $D_{2h}$
point group invariance~\cite{LVoon}, which agrees well with a tight binding (TB) model. ~\cite{XZhou4} The Hamiltonian is given by,~\cite{RFei,YJiang,Paulo,XZhou}
\begin{equation}
H=
\begin{bmatrix}
E_{c}+\alpha_{c} k^2_{x}+ \beta_{c} k^2_{y} & \gamma k_{x} \\
\gamma k_{x} & E_{v}-\alpha_{v} k^2_{x}- \beta_{v} k^2_{y}%
\end{bmatrix}+H_{v}.
\label{H}
\end{equation}
$H_{v}$ is the potential term. The band parameters are $\alpha _{c}=\hbar ^{2}/2m_{cx}$, $\beta _{c}=\hbar
^{2}/2m_{cy}$, $\alpha _{v}=\hbar ^{2}/2m_{vx}$, $\beta _{v}=\hbar
^{2}/2m_{vy}$. We list all the $\vec{k}\cdot\vec{p}$ parameters in table \ref{table}.

\begin{table}
\centering
\caption{Values of the $\vec{k}\cdot\vec{p}$ parameters}
\begin{tabular}{lllllllll}

\toprule
   $m_{cx}$ & $m_{cy}$ & $m_{vx}$ & $m_{vy}$  & $\alpha_{c}$  & $\beta_{c}$ & $\alpha_{v}$  & $\beta_{v}$ & $\gamma$ \\
\hline
   0.793  &  0.848   & 1.363   & 1.142 & 4.816 & 4.504 & 2.802   & 3.344  &-0.5231  \\

\botrule
  \end{tabular}
\footnotesize{$m_{cx}$-$m_{vy}$ are in the unit of electron mass $m_{e}$.}\\
\footnotesize{$\alpha_{c}$-$\beta_{v}$ are in the unit of $10^{-2}$eV $\cdot$ $nm^{2}$}\\
\footnotesize{$\gamma$ is in the unit of eV $\cdot$ nm}
  \label{table}
\end{table}

We use $k_{x}$($k_{y}$) to represent the wave vector along the armchair(zigzag) direction of phosphorene. Now we consider the arbitrary-direction periodic superlattice over the phosphorene induce periodic potential. 
For calculation convenience, we choose the new axis of the wave vector $k^{'}_{x}$($k^{'}_{y}$) to be vertical to periodic superlattice(parallel to the superlattice). The transformation relation between the vector is
\begin{equation}
\begin{split}
k_{x}^{'}=k_{x}cos\theta+k_{y}sin\theta \\
k_{y}^{'}=-k_{x}sin\theta+k_{y}cos\theta
\end{split}
\label{transformation}
\end{equation}
In the new coordination, the accordingly Hamiltonian can be written as,
\begin{widetext}
\begin{equation}
H=
\begin{bmatrix}
E_{c}+\alpha_{c}^{'} k^{'2}_{x}+ \beta_{c}^{'} k^{'2}_{y}+2\lambda_{c}k_{x}^{'}k_{y}^{'}-eU(x) & -\gamma sin\theta k_{y}^{'}+\gamma cos\theta k_{x}^{'} \\
-\gamma sin\theta k_{y}^{'}+\gamma cos\theta k_{x}^{'} & E_{v}-\alpha_{v}^{'} k^{'2}_{x}-\beta_{v}^{'} k^{'2}_{y}-2\lambda_{v}k_{x}^{'}k_{y}^{'}-eU(x)%
\end{bmatrix}.
\label{H}
\end{equation}
\end{widetext}
$\alpha_{c}^{'}$($\alpha_{v}^{'}$), $\beta_{c}^{'}$($\beta_{v}^{'}$) and $\lambda_{c}$($\lambda_{v}$) are new band parameters for electrons in the conduction band (valence band) as shown below:
\begin{equation}
\begin{split}
\alpha_{c,v}^{'}=\alpha_{c,v}cos^{2}\theta+\beta_{c,v}sin^{2}\theta \\
\beta_{c,v}^{'}=\alpha_{c,v}sin^{2}\theta+\beta_{c,v}cos^{2}\theta \\
\lambda_{c,v}=(-\alpha_{c,v}+\beta_{c,v})sin\theta cos\theta
\end{split}
\label{transformation}
\end{equation}
In the new coordination, the periodic potential $U(x)=V_{0}cos(2 \pi x/W)$, $W$ is the width of potential stripes, we denote $L$ as the periodic length of the superlattice superlattice as shown in Fig.~\ref{fig1}. $W=10$ nm, $L=20$ nm and $V_{0}=0.02eV$.

The envelope functions may be combined into a two-component spinor $\Psi
=(\Psi _{c},\Psi _{v})$ which satisfies a Dirac equation $H\Psi =E\Psi$. The
electron wave functions are expanded in a plane wave basis confined by the
large hard wall box. The wave function $\Psi $ for electron can be expanded
as
$\Psi(k_{x},k_{y})=\sum_{n}\mathbf{C}_{n}\phi_{n}(k_{x},k_{y})
=\sum_{n}1/\sqrt{L}\mathbf{C}_{n}e^{i(\frac{2n\pi x}{L}%
+k_{x}x)}e^{i k_{y}y},$
where $k_{x}$ ($k_{y}$) is the wave vector vertical to (parallel to) the external field, and the expansion coefficient $\mathbf{C}_{n}$ a two-component column vector. The wave function can be calculated numerically in the basis set with the
periodic boundary conditions.

The interaction Hamiltonian, $H_{int}=H(\vec{p}+e\vec{\mathcal{A}})-H(\vec{p}%
)$, describe the coupling between the Dirac fermion and the photon within
the electrical dipole approximation. For the $\sigma \pm$ circularly polarized lights, the vector potentials follow $\vec{\mathcal{A}}=(\mathcal{A}_{x}\pm i\mathcal{A}_{y}%
)e^{-i\omega t}$ respectively.
$|i>$ denotes the initial states in the lower cones that are hole or valence
like, $|f>$ denotes the final states in the upper cone states that are
electron of conduction like. The resulting optical transition rate of e-h pair
between valence and conduction band is $|<f|H_{int}|i>|$,
\begin{equation}
H_{int}=
\begin{bmatrix}
0 & \gamma e/ \hbar A_{x} &0  &0\\
\gamma e/ \hbar A_{x} & 0   &0   & 0 \\
0 &  0   &   0  & \gamma e/ \hbar A_{x}  \\
0   &  0  &  \gamma e/ \hbar A_{x} &  0
\end{bmatrix}.
\label{H_int}
\end{equation}  
In the inclined superlattice case, the off-diagonal element become $\gamma (isin \theta+cos \theta) e/ \hbar A_{x}$. 
The transition rate is given by,
$w_{if}=2\pi \delta (E_{f}-E_{i}-\hbar \omega )|<f|H_{int}|i>|^{2}$.
Finally we can obtain the optical absorption
rate by the integral of transition rates in $k$ space,
$\alpha (\hbar \omega )=\iint_{k_{x},k_{y}}\sum_{i,f}w_{if}dk_{x}dk_{y}.$

\section{Numerical results and discussions}

\begin{figure}[tbp]
\includegraphics[width=\columnwidth]{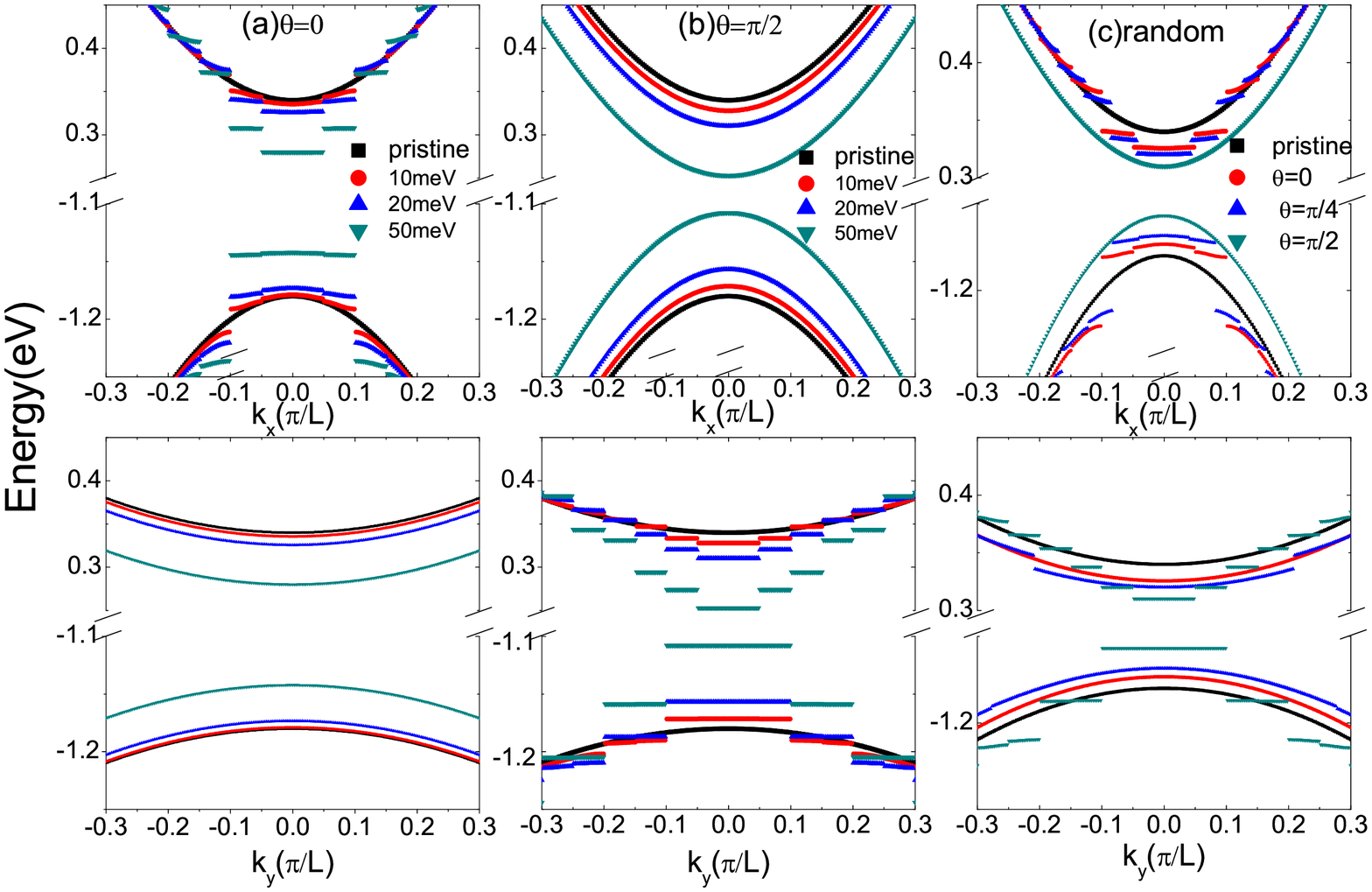}
\caption{The unfolded low energy dispersions of phosphorene  with different periodic potential strength along the (a) armchair and (b) zigzag direction. (c) The same as (a) and (b), but with different phosphorene metal stripe superlattice twisted angle $\theta$ and fixed potential strength $V=0.02$ eV. }
\label{band}
\end{figure}

\begin{figure}[tbp]
\includegraphics[width=\columnwidth]{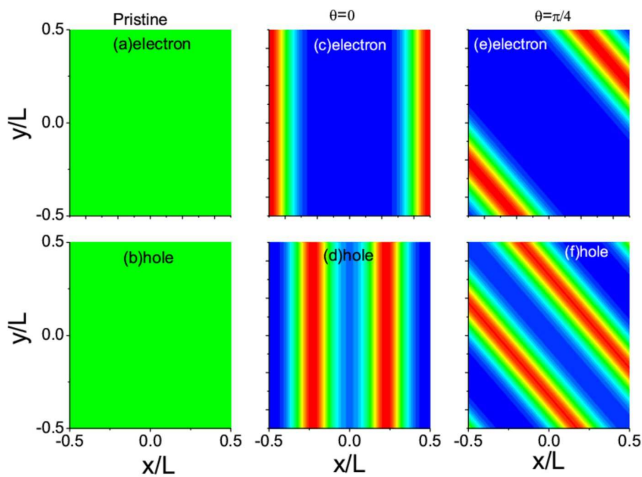}
\caption{The density distribution of electrons (holes) in the lowest conduction (highest valence) subband (a)-(b) with out potential superlattice, and (c)-(f) with potential superlattice of different twisted angle $\theta$.}
\label{density}
\end{figure}

We start by investigating the energy dispersion of phosphorene along the armchair (Fig.~\ref{band}(a)) and zigzag (Fig.~\ref{band}(b)) direction when the superlattice directions are aligned with the rectangular unit cell axes directions. In pristine phosphorene, the rectangular unit cell gives rise to anisotropic energy dispersion along the two orthogonal $k_{x}$ and $k_{y}$ directions, indicating the anisotropic group velocity and effective masses. In the presence of an external periodic potential field along the armchair direction, we can find that the band gap is opened at the Brillouin zone boundary of the superlattice, and becomes larger as the strength of potential increase. It arises from the periodic perturbation effect at a reduced Brillouin zone boundary, e.g. $k_{x}(k_{||})=\pm\pi/L+2n\pi/L$. In the transverse direction $E[k_{y}(k_{T})]$ when $\theta=0$, the energy spectrum does not split, since the external potential is constant. The band gap decrease with increasing the external field strenth, which is consistent with previous studies of phosphorene superlattice with electric potentials~\cite{Ono}. The external field modulation is more significant in $y$ direction (i.e., $\theta=\pi/2$) superlattice than that in $x$ direction ( $\theta=0$) superlattice.

Then we focus on the energy dispersion of a phosphorene with external periodic potential in arbitrary direction in Fig.~\ref{band}(c). Compare to the pristine phosphorene in black in Fig.~\ref{band}(c), the energy dispersions of phosphorene with periodic potential modulations with rotated angle, i.e., $\theta=0$, $\theta=\pi/4$ and $\theta=\pi/2$ are shown in Fig.~\ref{band}(c) with fixed strength $V=20 meV$. We find strongly superlattice orientation dependent energy dispersions, bandgap variations, energy level (bond states) formations. Due to the smaller slope of energy spectrum as respect to $k_{y}$, The electrons have larger effective mass and smaller kinetic energies, so it is more pronounced response to the periodic potential perturbations. It is worthy note that, the band splitting is pronounced along the periodic direction of external field along which the energies is effectively modulated, while the energy dispersion along the transversal direction is almost immune to external fields. Notably, flat subbands are formed in the presence of external superlattice in arbitrary direction. Flat subbands indicate that boundstates are confined in each potential well. To confirm the prediction of carrier localization by energy dispersion analysis, we plot the density distribution of both electrons and holes in the lowest conduction subband and highest valence band respectively in Fig.~\ref{density}. Not only the localized carrier distributions are observed, but also the spacially separation of electrons and holes are shown. Because the Sinusoidal $U$ in Eq.~\ref{H} leads to spatially shifted potential well of electrons and holes by $L/2$. For a preliminary summary, the proposed superlattice can effectively tune the anisotropy of energy dispersion and density distribution. Accordingly, band engineering enables the modulations of carrier effective mass and wavefunctions coupling as we discussed in the following.

\begin{figure}[tbp]
\includegraphics[width=0.95\columnwidth]{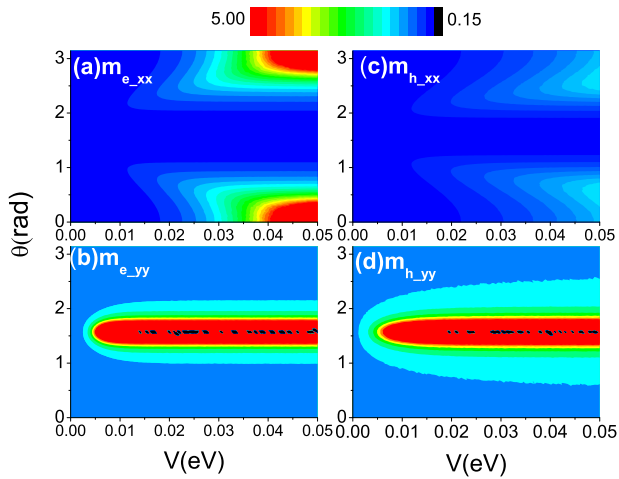}
\caption{The contour plot of anisotropic effective mass of  (a)-(b) electron and (c)-(d) hole as a function of the superlattice periodic potential strength $V$ and twisted angle $\theta$. }
\label{mass}
\end{figure}

The modulated anisotropy can be quantitatively analyzed by monitoring the effective mass tensor. Fig.\ref{mass} shows the contour plot of effective masses components along the two crystalline directions of phosphorene for both electron and holes as functions of the superlattice rotated angle $\theta$ and the external potential strength $V$. With fixed direction of $\theta=0$ (along x direction), we can find the effective mass of electron along the $x$ direction increases when the external potential increases (see Fig.\ref{mass}(a)), corresponding to the the flat energy band tuned by the metal stripe superlattice (see Fig.\ref{band}). When the $\theta$ increases, the effective mass along $k_{x}$ decreases. Finally when the $\theta=\pi/2$, it plays a negligible effect on the energy dispersion along $k_{x}$, so as the effective mass. The effective mass of electron along the $y$ direction is opposite as shown in Fig.\ref{mass}(b) . When $\theta=\pi/2$ (the $y$ direction superlattice), increasing the potential, we can find the increasing of the effective mass (flat band). But the potential have no effect on the effective mass when the $\theta=0$ (the $x$ direction superlattice). We also can find the potential along the $y$ direction has more significant impact on the energy dispersion and effective mass. The effective mass and the anisotropy of holes can also be effectively modulated due to the same mechanism (see Figs.\ref{mass}(c) and (d)).

\begin{figure}[tbp]
\includegraphics[width=0.95\columnwidth]{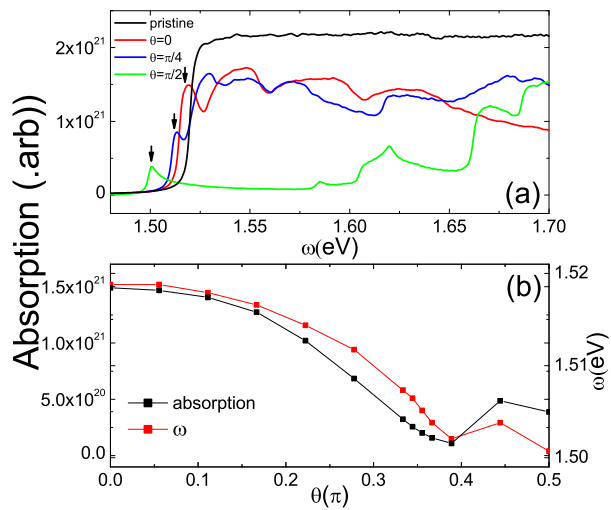}
\caption{(a) The optical absorption spectrum as function of photon energy with the phosphorene metal stripe superlattice twisted angle. (b) The optical absorption and the electron-hole transition energy as function of the superlattice twisted angle. }
\label{pl}
\end{figure}

To develop the potential applications in optoelectronics, we investigate the superlattice orientation dependant optical absorption rate of the phosphorene superlattice. Because the periodic potential superlattice has the equivalent effect on the clockwise and anti-clockwise ($\sigma \pm$ ) absorption rate, we only plot the $\sigma+$ absorption rate in Fig.\ref{pl}. We use broadening
factor of $0.15$ $meV$ to smoothen the absorption spectrum. In Fig.\ref{pl}, the optical
absorption spectrum indicates useful band structure information guaranteed
by the selection rule expressed as $\delta (E_{f}-E_{i}-\hbar\omega)$. In the presence of a potential superlattice, the maximum optical absorption decreases apparently due to the suppression of the wavefunction overlap between the CBM electrons and the VBM holes, which is indicated by the spatially separated charge density distribution as we discussed in Fig.~\ref{density}. In Fig.~\ref{pl}(a), different configurations of
periodic potentials show different optical absorption spectra. For a $x$ direction (armchair direction) potential superlattice, the optical absorption spectrum with many
wide step indicates band splitting arising from the electric potential modulations as we discussed before. When the potential superlattice rotates from the armchair direction by an angle $\theta$, we can find the shifting of band edge to lower energies (see arrows in Fig.~\ref{pl}(a)). When the rotated angel of the superlattice $\theta=\pi/2$, the bandgap reaches the minimum, which is consistent with the variations of electronic structures. The strength of absorption also decreases with increasing $\theta$. It relates to the decreasing of the carrier density. For a $y$ direction (zigzag direction) potential superlattice, we can find a single absorption peak in the band edge and more oscillations when increasing the energy
(frequency) of the incident light due to larger band splitting by the superlattice modulations. To demonstrate the trend clearly, we plot the absorption rate and electron-hole pair transition energy of incident light as function of superlattice rotated angle $\theta$ in Fig. \ref{pl}(b). We can see the decreasing trend in both curves. There are fluctuation humps around $\theta=0.4\pi$. This feature arises from the change of band splitting as respect with $k_{x}$ to $k_{y}$. The distinct optical absorption
spectra provide an effective way to detect the anisotropic energy properties of superlattice with different periodic orientations. The phosphorene
superlattice is a promising platform for potential application in anisotropic optoelectronic devices.

\section{Conclusions}

In this work, we theoretically investigate the electronic and optical
properties of phosphorene artificial superlattice utilizing the $\vec{k}\cdot\vec{p}$ method. Combined with the crystalline induced anisotropy, the
potential fields make additional subtle tuning of intrinsic crystal symmetry and develop an effective way to modulate the electrostatics and optoelectronic properties of phosphorene superlattice. We demonstrated that the anisotropic energy dispersions can be tuned by the potential superlattice configurations. The potential fields and periodic orientations proposed in this work play important roles in determining the anisotropic effective mass and optical absorption spectrum. This work sheds new light on potential applications of optoelectronic devices based on the phosphorene in rotatable periodic potentials.

\begin{acknowledgments}
This work was supported by the NSFC (Grant No. 61774168, 11774021, 61974027), the MOST of China (Grants No. 2016YFA0202300, 2017YFA0303400, 2014CB848700), and the Opening Project of MEDIT, CAS.
\end{acknowledgments}

\end{document}